\documentclass{article}
\usepackage{./_config/spconf}
\usepackage{amsmath}
\usepackage{graphicx}
\usepackage[colorlinks=true,
            linkcolor=black,
            filecolor=magenta,      
            urlcolor=blue,
            citecolor=black]{hyperref}

\newcommand{\etal}{\emph{et al.~}}
\newcommand{\comment}[1]{}
\title{G-PCC post-processing using fractional super-resolution}
\name{Renan U.B.Ferreira, Tomás M. Borges, Diogo, C. Garcia, Ricardo L. de Queiroz%
\thanks{\scriptsize{Work partially supported by CNPq under grants 88887.600000/2021-00 and 301647/2018-6.}}}
\address{University of Brasilia\\
\normalsize \emph{renan@unb.br, tomas@divp.org, diogogarcia@unb.br and queiroz@ieee.org}}
\begin{document}
%
\maketitle
\begin{abstract}
We present a method for post-processing point clouds' geometric information by applying a previously proposed fractional super-resolution technique to clouds compressed and decoded with MPEG's G-PCC codec. In some sense, this is a continuation of that previous work, which requires only a down-scaled point cloud and a scaling factor, both of which are provided by the G-PCC codec. For non-solid point clouds, an \textit{a priori} down-scaling is required for improved efficiency. The method is compared to the GPCC itself, as well as machine-learning-based techniques. Results show a great improvement in quality over GPCC and comparable performance to the latter techniques, with the advantage of not needing any kind of previous training.

\end{abstract}
\section{Introduction}
\label{sec:intro}

Point clouds are sparse 3D signals composed by geometry and attributes information (which may include color, reflectance, normal vectors, etc.), and have been in the spotlight of researchers in recent years for its usability in applications such as AR/VR (augmented and virtual reality) \cite{Schwarz2019REALTIMEDA,mpeg:m49235}, telecommunications \cite{mpeg:m38136}, autonomous vehicle \cite{mpeg:m43647} and world heritage \cite{mpeg:m37240}. 
Because of the need for compression for either storage or transmission, the Motion Picture Experts Group (MPEG) has been directing efforts for the compression of point clouds in two fronts, i.e., geometry-basde point cloud compression (G-PCC) and video-based point cloud compression (V-PCC). 
Concerning the geometric information, each point may be expressed as a list $V$ of unordered ternary information, such that the $n$-th point is $\textbf{v}_n = (x_n, y_n, z_n)$. As for its coding, while V-PCC uses video codecs for encoding its projection in a plane, G-PCC uses the \emph{octree} structure \cite{article:danilo:2020}.

Geometry coding under G-PCC may be either lossless or lossy. To encode geometry in a lossy way, G-PCC prunes the octree in order to eliminate nodes after a certain pre-determined level. 
In this process, G-PCC uses a coordinate transformation at the encoder, which down-scales the original point cloud geometry $V$ such that, for the $n$-th point of $V_d$
\begin{equation}
    \textbf{v}_{d_n}=\texttt{round}\left(\frac{\textbf{v}_n-T}{s}\right),
    \label{eq:downsampling}
\end{equation}
where, $\texttt{round}(\cdot)$ is the function that rounds the components of a vector to the nearest integer,
\begin{equation}
    T=(\min x_n, \min y_n, \min z_n),
\end{equation}
and $s>1$ is the scale factor, given by the inverse of the \texttt{positionQuantizationScale} parameter \cite{mpeg:codec_gpcc}.
This scaling reduces the number of points to be encoded due to rounding and to duplicate point removal, making $V_d$ a coarser geometry when compared to $V$. The larger the scale factor $s$, the lower the encoding rate, and consequently, the lower the number of output points, and the coarser the geometry.
At the decoder side, the scaled geometry is expanded and shifted back to its original position using the same values of $s$ and $T$.

\begin{figure*}[t]
\centering
\begin{minipage}[b]{0.33\linewidth}
  \centering
  \centerline{\includegraphics[width=\textwidth]{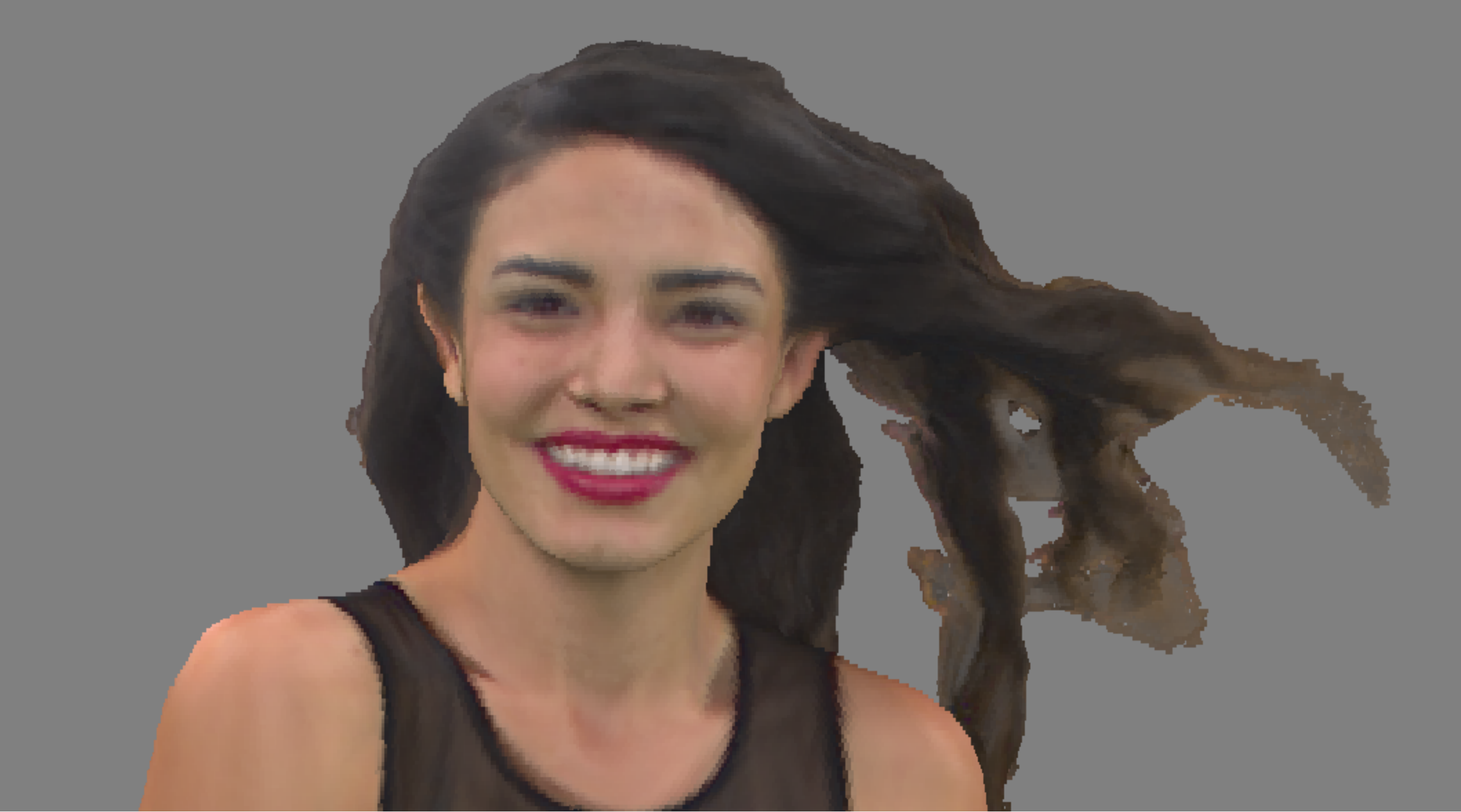}}
  \centerline{(a)}\medskip
\end{minipage}
\begin{minipage}[b]{0.33\linewidth}
  \centering
  \centerline{\includegraphics[width=\textwidth]{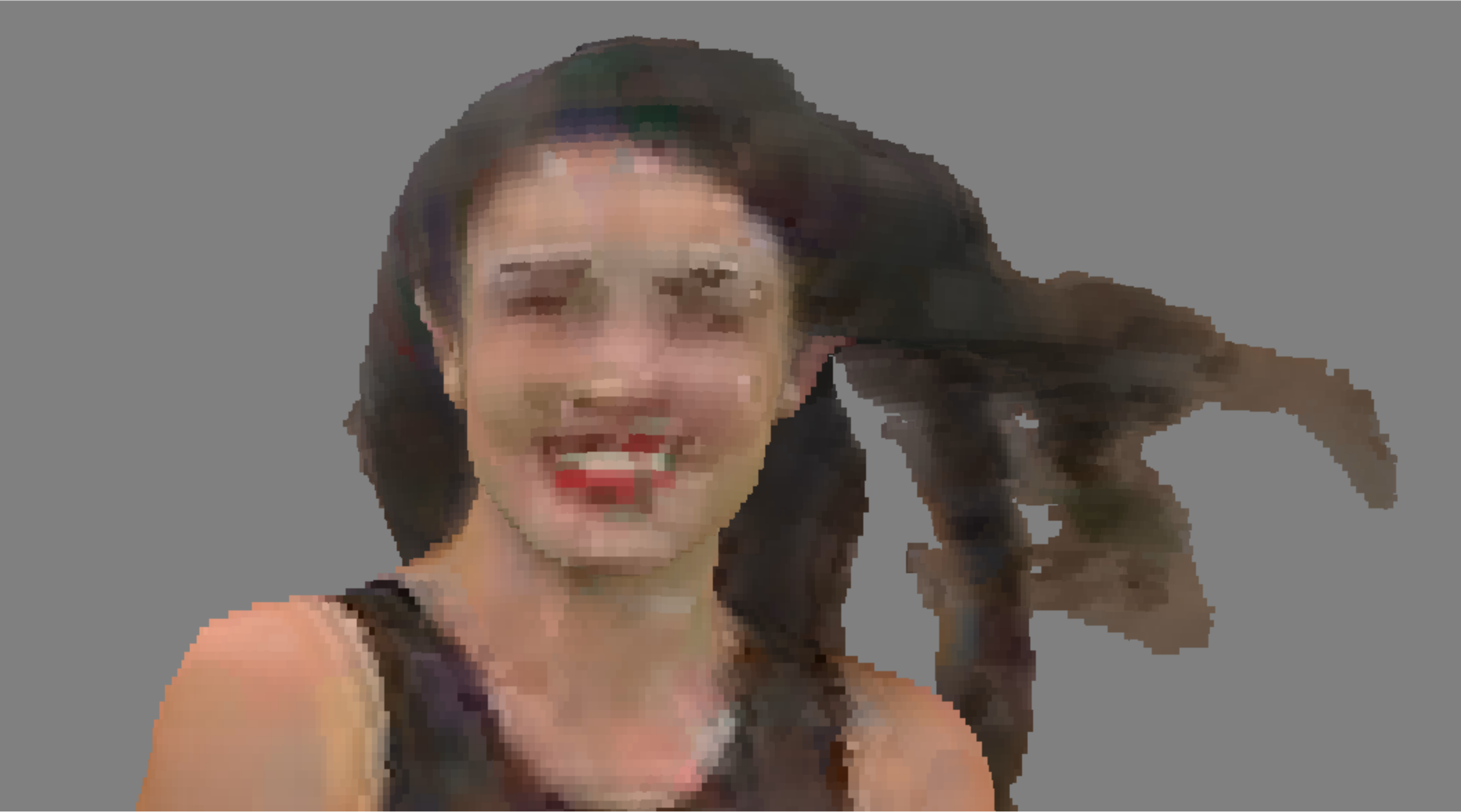}}
  \centerline{(b)}\medskip
\end{minipage}
\begin{minipage}[b]{0.33\linewidth}
  \centering
  \centerline{\includegraphics[width=\textwidth]{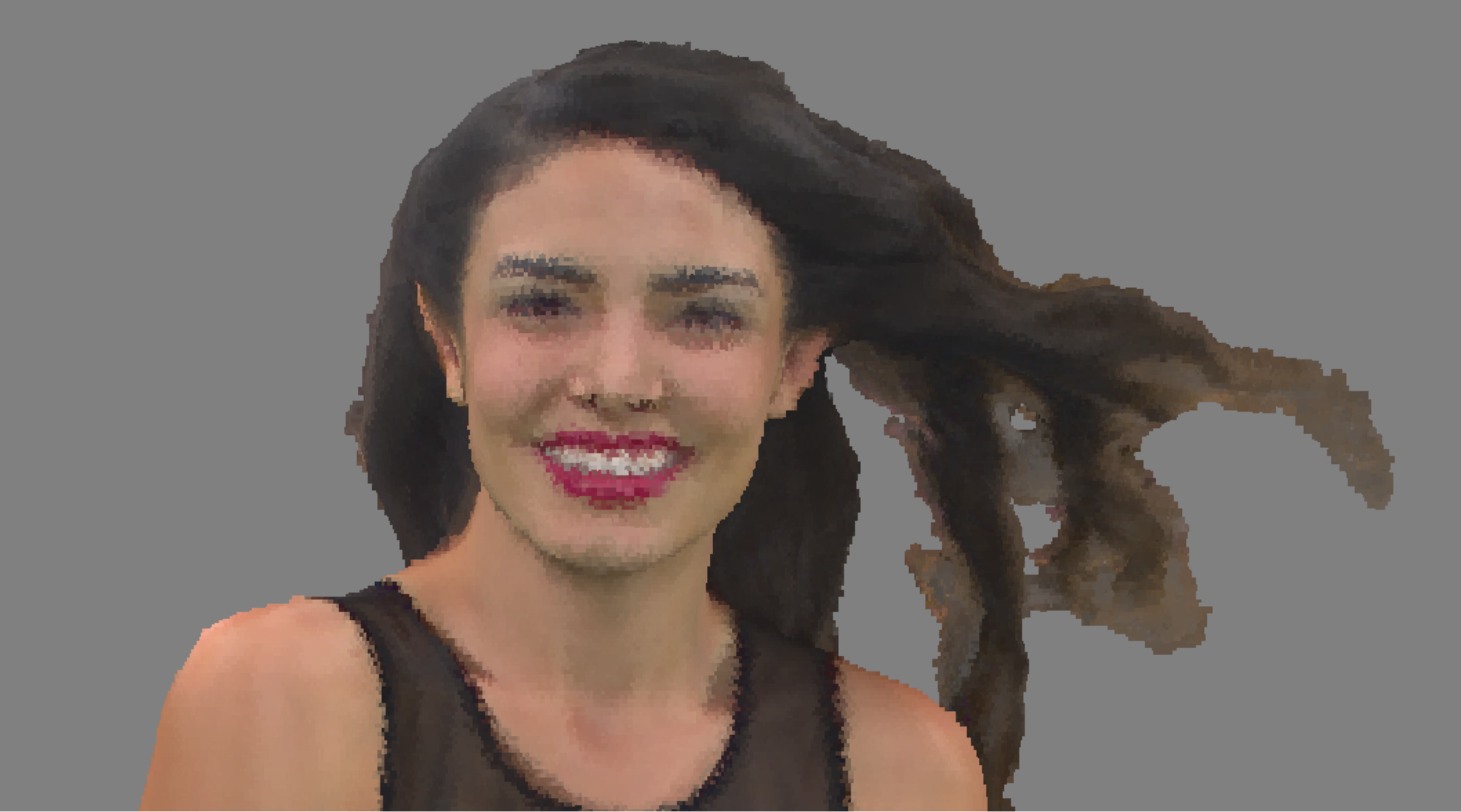}}
  \centerline{(c)}\medskip
\end{minipage}

\caption{Example of point clouds in the post-processing method: (a) $V$; (b) $V_{d}$ with NNI; (c) $V_{sr}$.}
\label{fig:method_example}
\end{figure*}

Some previous techniques have been proposed in order to improve coding efficiency for these lossy cases, such as the use of slicing interpolation \cite{mpeg:m42519} inside G-PCC or even completely new approaches, such as using dyadic decomposition \cite{inproc:Eduardo}.
On the other hand, the use of a lookup table (LUT) based on neighborhood inheritance has been used for context generation on geometry coding of dynamic voxelized point clouds \cite{article:diogo:2020} as a replacement to the octree. 
Most recently, Borges \etal \cite{article:tomas:2022} proposed the use of LUTs relating the downsampled neighbourhood of a given voxel with its children occupancy for super-resolution (SR) 
of voxelized point clouds downsampled at arbitrary fractional scales.
Assuming that self-similarities are somewhat maintained at different scales, and taking into account the irregulaties of fractional downsampling, they estimate children occupancy based on the neighborhood of downsampled versions of a single input point cloud. 
Results have shown that the fractional SR provides better results than other solutions, such as using nearest-neighbours interpolation (NNI) and NNI followed by laplacian smooting \cite{inproc:LS}.
Figure \ref{fig:method_example} shows an example of such results.
As a continuation of that work, we apply the fractional SR technique as a post-processing tool for point clouds decoded with G-PCC.

\section{Post-Processing Method}
\label{sec:format}

Since the fractional SR technique of Borges \etal \cite{article:tomas:2022} can be employed for arbitrary scale factors, it can be used to improve quality of point clouds downsampled using Eq. \eqref{eq:downsampling}, provided that $s$ is known.
Thus, it can be used to post-process point clouds geometric information, after being encoded and decoded using the G-PCC codec in lossy-geometry configuration.

\begin{figure}[htb]
    \centering
    \includegraphics[scale=1]{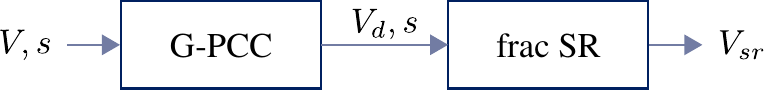}
%
    \caption{The fractional SR is applied to the decoded point cloud using the same scale factor as used in the downsampling of the G-PCC codec.}
    \label{fig:method}
\end{figure}

Recognizing that the lossy encoding of geometry in G-PCC is but a downsampling, one can improve the downsampled point cloud geometry $V_d$ using the fractional scale factor $1<s\leq 2$ from the encoder, to generate $V_{sr}$.
Figure \ref{fig:method} shows the general idea of the proposed post-processing method. For some coding conditions, the value of $s$ may be greater than 2. In this situations, the fractional SR may be applied successively.

\begin{figure}[htb]
    \centering
    \includegraphics[scale=1]{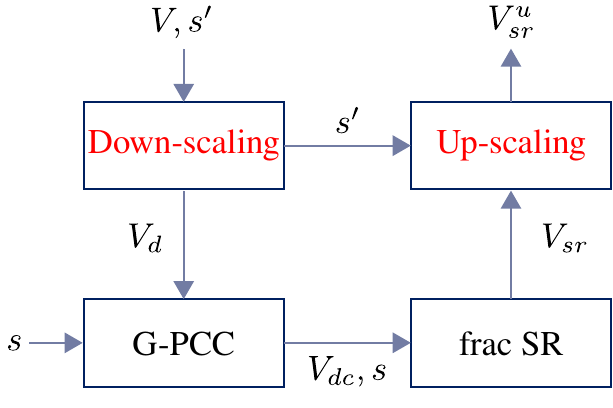}
%
    \caption{For sparse clouds, a down-scaling is applied previously to the encoding a a up-scaling is applied after the SR process.}
    \label{fig:dus}
\end{figure}

The SR method relies on the use of the adjacent neighbourhood of a given voxel in order to predict its children nodes. For some sparser point clouds, the immediate neighbours may not be available for some (or maybe all) voxels. In order to surpass this problem, we down-scale the input point cloud by a factor $s'$ prior to the encoding. Then, a subsequent up-scaling of the cloud $V_{sr}$ by the same factor is necessary, after the ``frac SR'' step, as depicted in Fig. \ref{fig:dus}. 

\section{Experimental Results}
\label{sec:pagestyle}

\begin{figure}[htb]
\centering
\begin{minipage}[t]{\linewidth}
  \centering
  \centerline{\includegraphics[width=7.5cm]{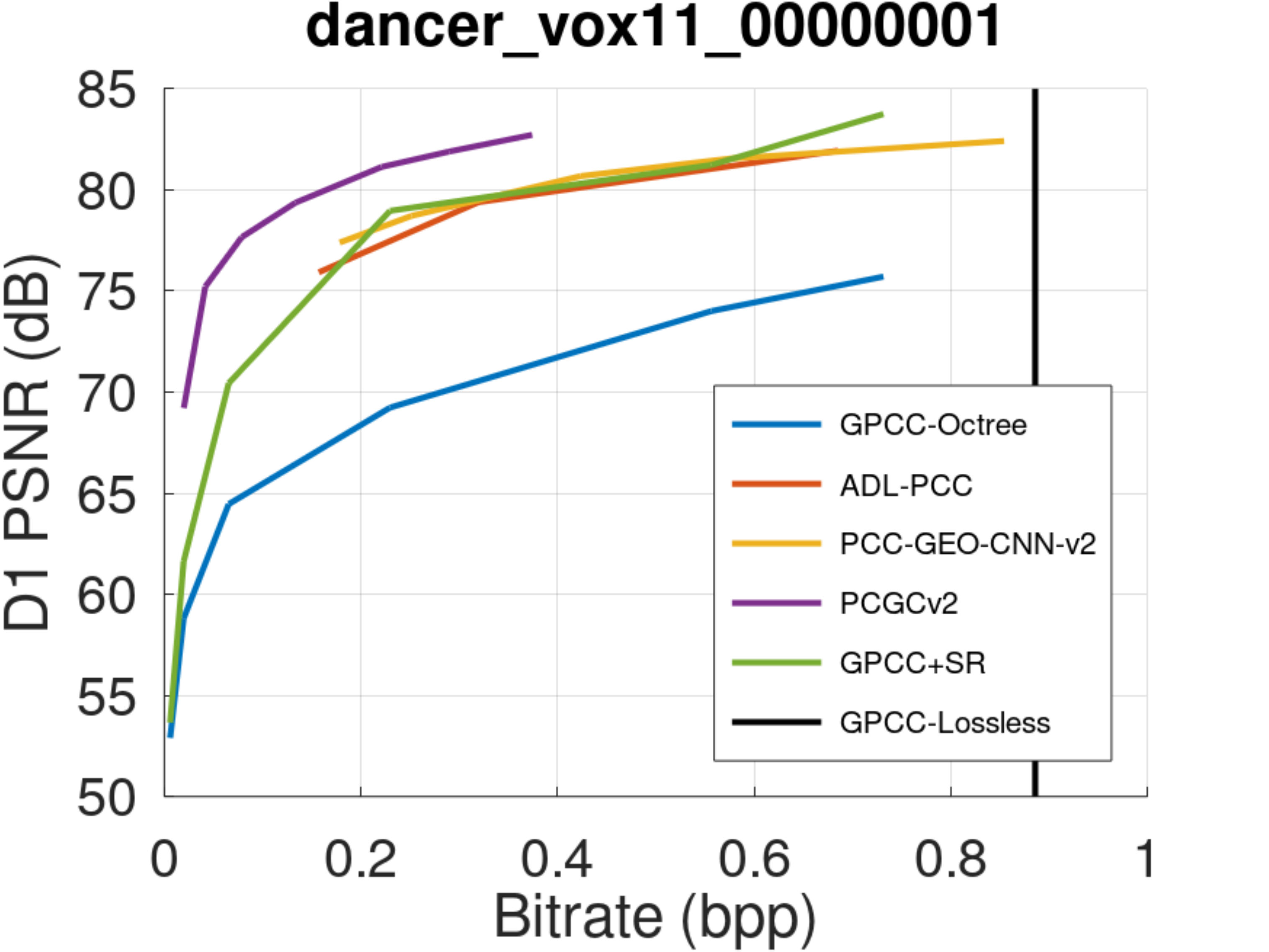}}
  \centerline{(a)}
  \medskip
\end{minipage}
\begin{minipage}[t]{\linewidth}
  \centering
  \centerline{\includegraphics[width=7.5cm]{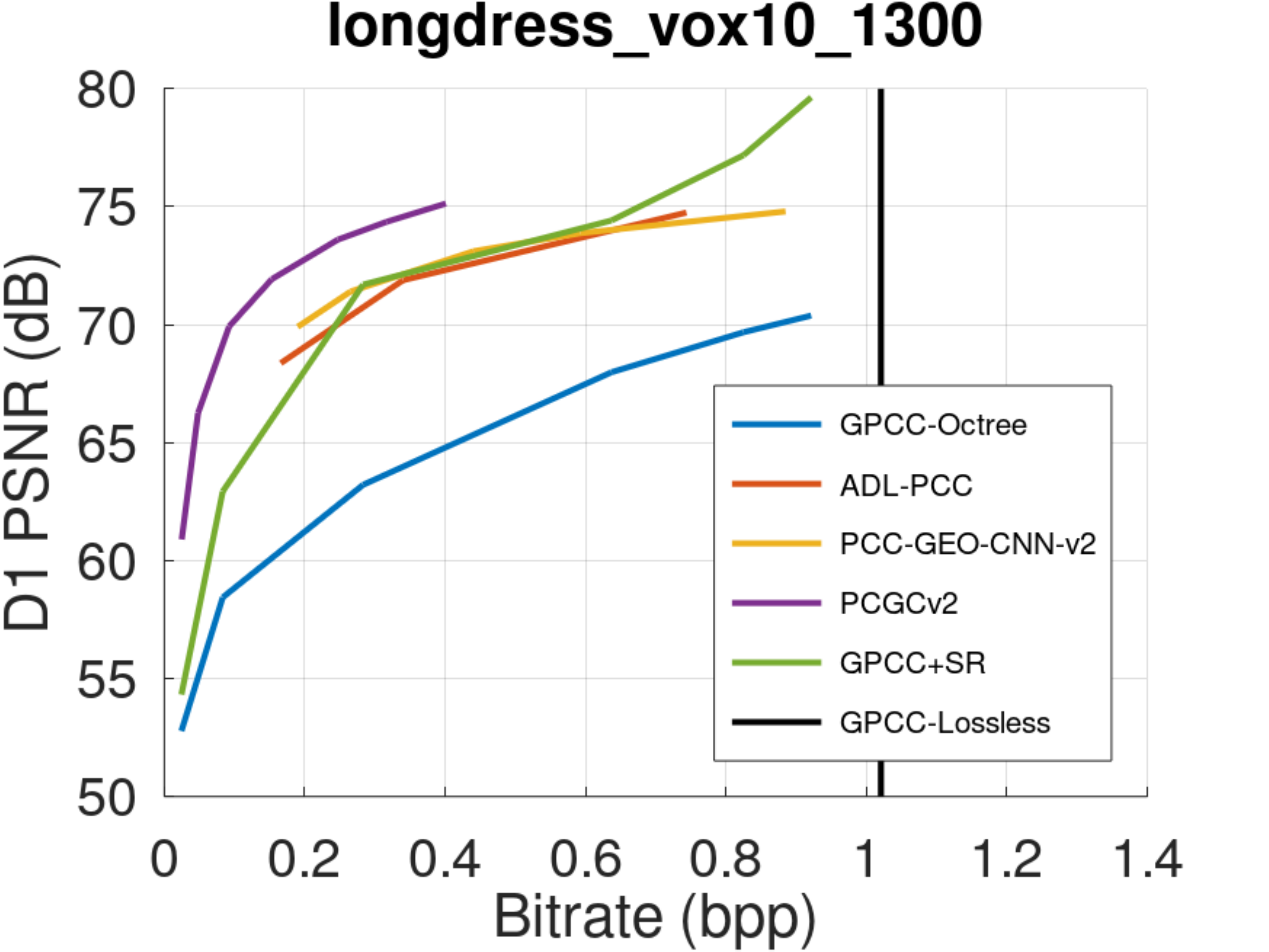}}
  \centerline{(b)}
  \medskip
\end{minipage}

\caption{Comparative results under D1 PSNR metric for solid point clouds: (a) \textit{dancer}, (b) \textit{longdress}.}
\label{fig:rd_curves1}
\end{figure}

\begin{figure}[htb]

\begin{minipage}[b]{\linewidth}
  \centering
  \centerline{\includegraphics[width=7.5cm]{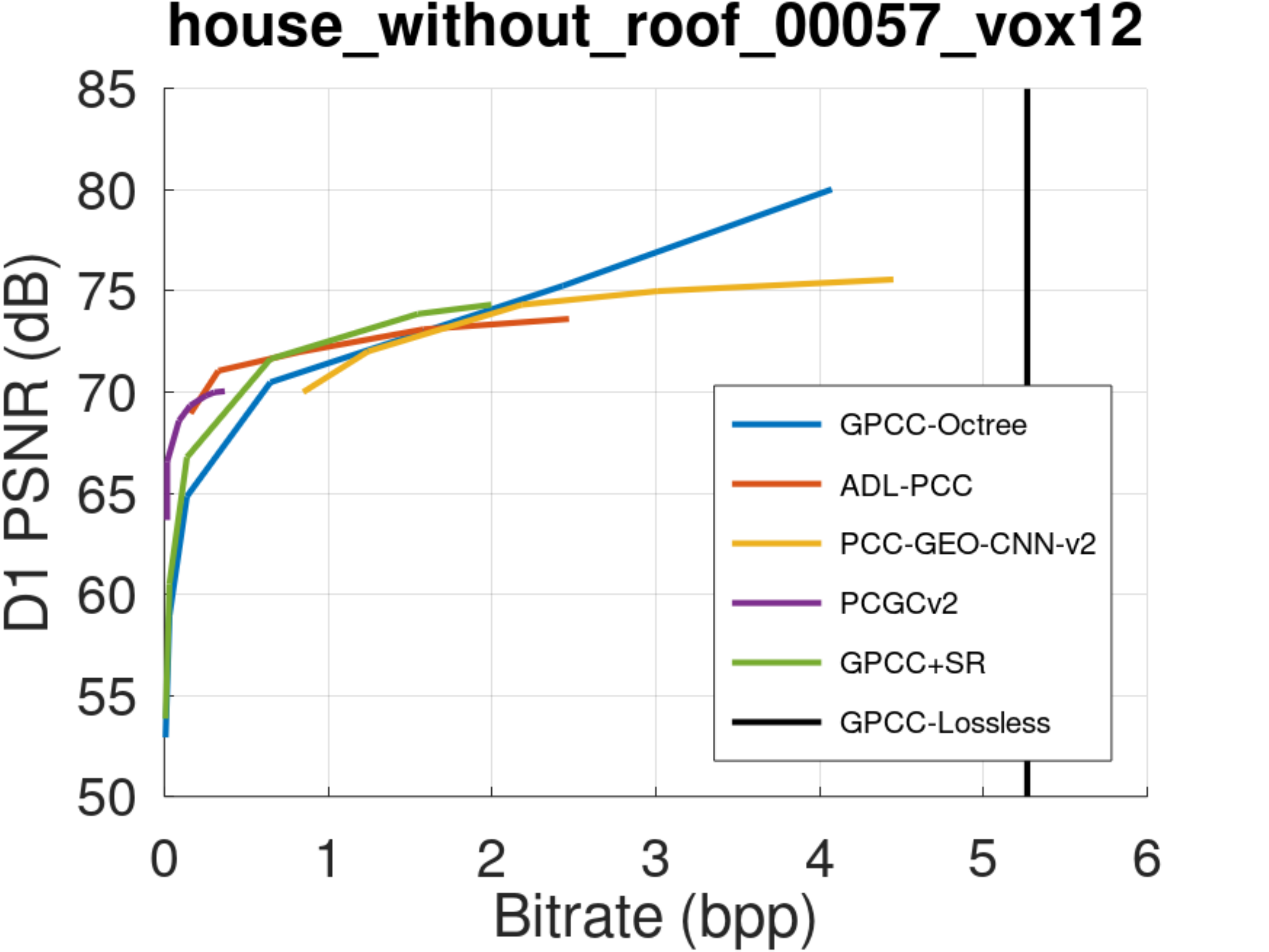}}
  \centerline{(a)}\medskip
\end{minipage}
\begin{minipage}[b]{\linewidth}
  \centering
  \centerline{\includegraphics[width=7.5cm]{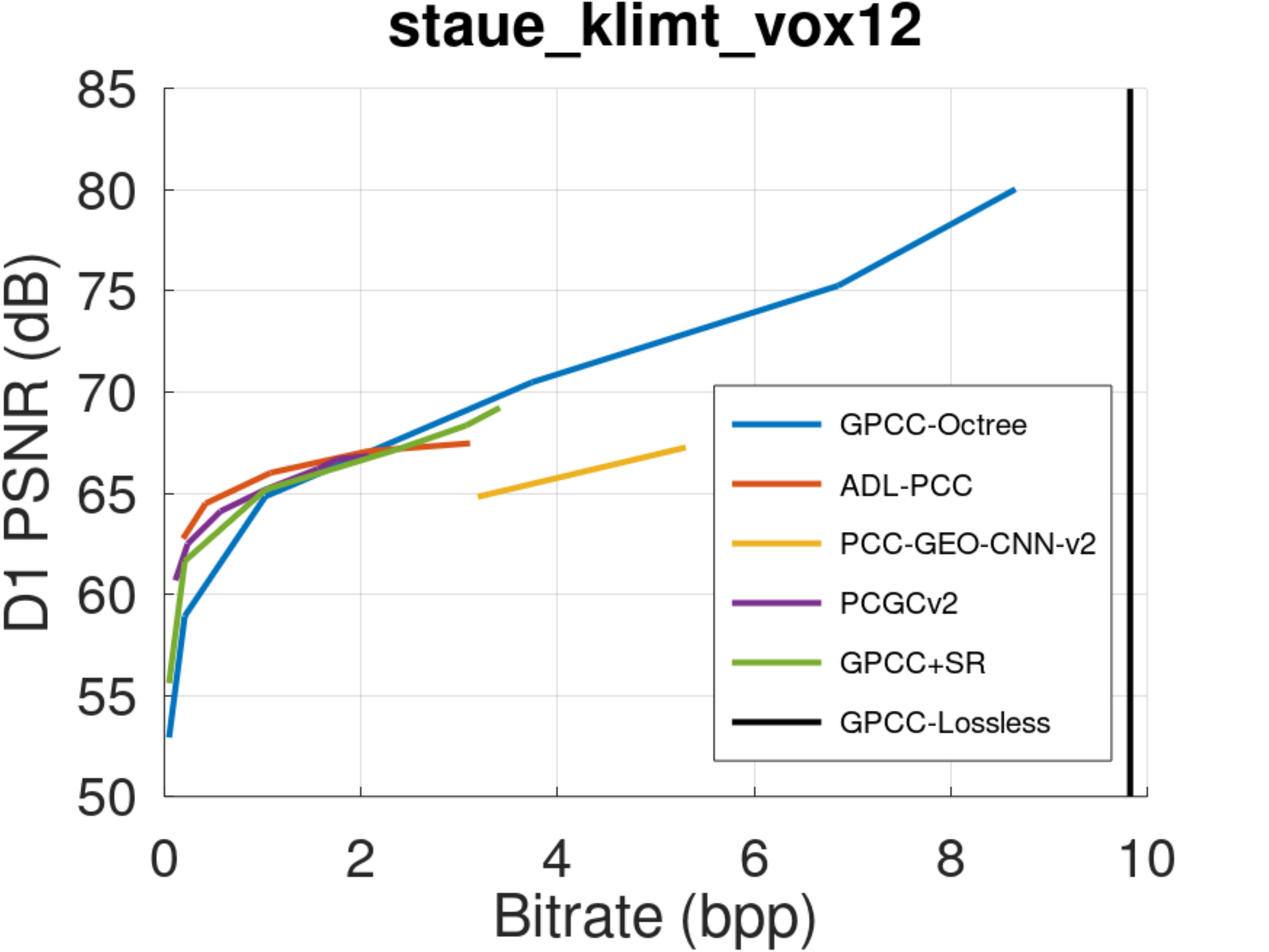}}
  \centerline{(b)}\medskip
\end{minipage}

\caption{Comparative results under D1 PSNR metric for non-solid point clouds: (a) \textit{house\_without\_roof}, (b) \textit{staue\_klimt}.}
\label{fig:rd_curves2}
\end{figure}

In order to test the post-processing method, we propose two distinct experiments, according to the sparsity of the input point clouds, for solid (voxelized point clouds with continuous surface) and non-solid clouds, according to the categories used in MPEG \cite{mpeg:m55485}. 
In the first one, we apply the fractional SR directly to the decoded versions of solid point clouds (as in Fig. \ref{fig:method}). The scaling factor is defined by the MPEG G-PCC's Common Test Conditions (CTC) \cite{mpeg:ctc_gpcc} according to the point cloud's geometry for each rate point to be tested, as shown in Table \ref{tab:rCTC}.
The greater the rate id (as in ``R6''), the greater the bit rate, the lower the value of $s$.

\begin{table}[htb]
\small
\centering
\begin{tabular}{||c c c c c c c||} 
 \hline
 Geometry precision & R6 & R5 & R4 &  R3 &  R2 &  R1  \\ [0.5ex] 
 \hline\hline
10 & 15/16 & 7/8 & 3/4 & 1/2 & 1/4 & 1/8 \\
11 & 7/8 & 3/4 & 1/2 & 1/4 & 1/8 & 1/16\\
 \hline
\end{tabular}
\caption{The \texttt{positionQuantizationScale} parameter used in G-PCC's CTC for different geometry precision to generate each rate point.}
\label{tab:rCTC}
\end{table}

In the second experiment, for testing non-solid point clouds, we apply a downscaling prior to encoding. The scaling factor $s'$ is defined as the highest power of 2 that makes $V_d$ keep approximately the same number of points as $V$, i.e., we enforce a downscaling only to densify the input geometry, avoiding over-decimating its points. The values are shown in Table \ref{tab:rCTC2}, for the two tested clouds. According to the G-PCC CTC's ``lossy-geometry'' section, the largest possible value used for this scaling would be 2048 (defined for point clouds with 20 bits in geometry precision). This means we only need 4 bits to transmit $s'$ as side information. Since the bitrate is calculated in bits per input points, i.e., the number of points in the original point cloud (which vary from hundreds of thousands to millions), these 4 bits are negligible.

\begin{table}[htb]
\small
\centering
\begin{tabular}{||c c||} 
 \hline
 Point cloud & $s'$ \\ [0.5ex] 
 \hline\hline
 house\_without\_roof\_00057\_vox12 & 2\\
 statue\_klimt\_vox12 & 4 \\
 \hline
\end{tabular}
\caption{Down-scaling factor for non-solid clouds.}
\label{tab:rCTC2}
\end{table}

For both experiments, we compare our results to those of G-PCC as well as to machine learning (ML) based end-to-end compression techniques \cite{quach2020improved,wang2020multiscale,article:AIPCC:ADL}, with test results provided by Zaghetto \cite{mpeg:zaghetto:AI1}. It is important to note that we only bring the results for the sequences used in the ML tests, for a direct comparison. Some of the clouds in the G-PCC's CTC where used in the training of some of these models, and therefore could not be used for testing.

Figures \ref{fig:rd_curves1} and \ref{fig:rd_curves2} show the rate-distortion curves for some of the obtained results, for solid and non-solid clouds, respectively. The rate is measured in bits per input point (the number of points in the original input point cloud), while the distortion is measured with the point-to-point PSNR (D1 PSNR) metric \cite{Girardeau2005}.

We note that our results bring great improvement over G-PCC, specially in the higher rates region. We also note that they are comparable to those of ML-based techniques, except for PCGCv2, which outperforms all others. However, we state the fact that our method doesn't require any type of training, while all the ML techniques require the training of models for each rate point shown in the curves.

We also notice a drop in the lower rates region of the curves. 
At this region, the scale factor $s$ is greater than 2, requiring to use the fractional SR more than once. This is not ideal because every time we apply it, we imply the neighbourhood for the next application, which may propagate errors.  

\begin{table*}[t]
\small
\centering
\begin{tabular}{||l c c c c c||} 
 \hline
 Cloud name & ADL-PCC & PCC-GEO-CNN-v2 & PCGCv2 & frac SR & frac SR (high-rates) \\ [0.5ex] 
 \hline\hline
 dancer\_vox11\_00000001 \cite{Owlii2017} & -77.04 & -87.81 & \textit{-93.63} & -59.79 & \textbf{-91.77} \\
 longdress\_vox10\_1300 \cite{8iDataset} &  -75.67   & -77.06 & \textit{-89.09} & -62.39 & \textbf{-80.30} \\
 loot\_vox10\_1200 \cite{8iDataset}  & -75.86 &  -74.88  & \textit{-90.38} & -64.40 & \textbf{-81.17} \\
 queen\_0200 & -75.93 & -79.46 & \textit{-92.85} & -70.23 & \textbf{-85.29} \\
 redandblack\_vox10\_1550 \cite{8iDataset}  &  -74.11  & \textbf{-75.88} & \textit{-88.62} & -62.52 & -73.89 \\
 soldier\_vox10\_0690 \cite{8iDataset}  & -76.99 & -75.30 & \textit{-89.39} & -65.63 & \textbf{-80.66} \\
 \hline
\end{tabular}
\caption{BR-Rate comparison to G-PCC for solid clouds, in rate \%.}
\label{tab:solid}
\end{table*}

\begin{table*}[t]
\small
\centering
\begin{tabular}{||l c c c c||} 
 \hline
 Cloud name  & ADL-PCC & PCC-GEO-CNN-v2 & PCGCv2 & frac SR+DUS\\ [0.5ex] 
 \hline\hline
 house\_without\_roof\_00057\_vox12 & \textbf{-41.14} &  26.91  & \textit{-87.77} & -30.84 \\
 statue\_klimt\_vox12 & \textit{-41.16} &  207.18  & \textbf{-39.50} & -37.80 \\
 \hline
\end{tabular}
\caption{BR-Rate comparison to G-PCC for non-solid clouds, in rate \%.}
\label{tab:sparse}
\end{table*}

In Tables \ref{tab:solid} and \ref{tab:sparse} we bring the BD-rate comparison of all results to that of G-PCC for solid and non-solid clouds, respectively. In Table \ref{tab:solid}, we also show a column comparing our results considering only the four higher rate points (``frac SR (high-rates)''), i.e. the four rightmost points from the graphs shown in Figure \ref{fig:rd_curves1}, where the rate ranges are closer to most of those from the ML techniques, bringing therefore a more adequate for comparison.
Table \ref{tab:sparse} shows ``frac SR+DUS'' (down-up-scaling) to highlight the difference between the two experiments.

\begin{figure}[!h]

\begin{minipage}[b]{0.48\linewidth}
  \centering
  \centerline{\includegraphics[width=0.95\linewidth]{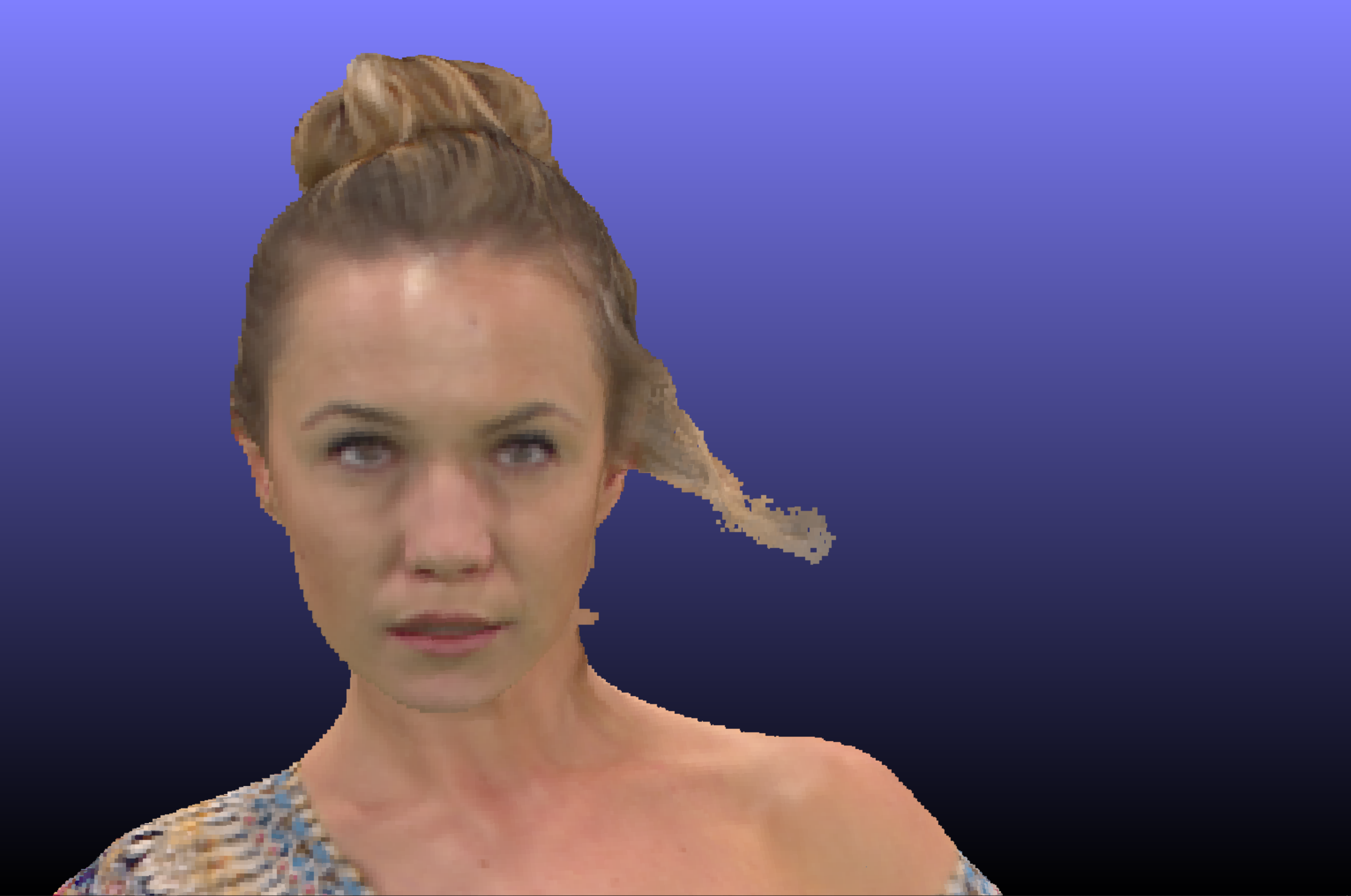}}
  \centerline{(a)}\medskip
\end{minipage}
\begin{minipage}[b]{0.48\linewidth}
  \centering
  \centerline{\includegraphics[width=0.95\linewidth]{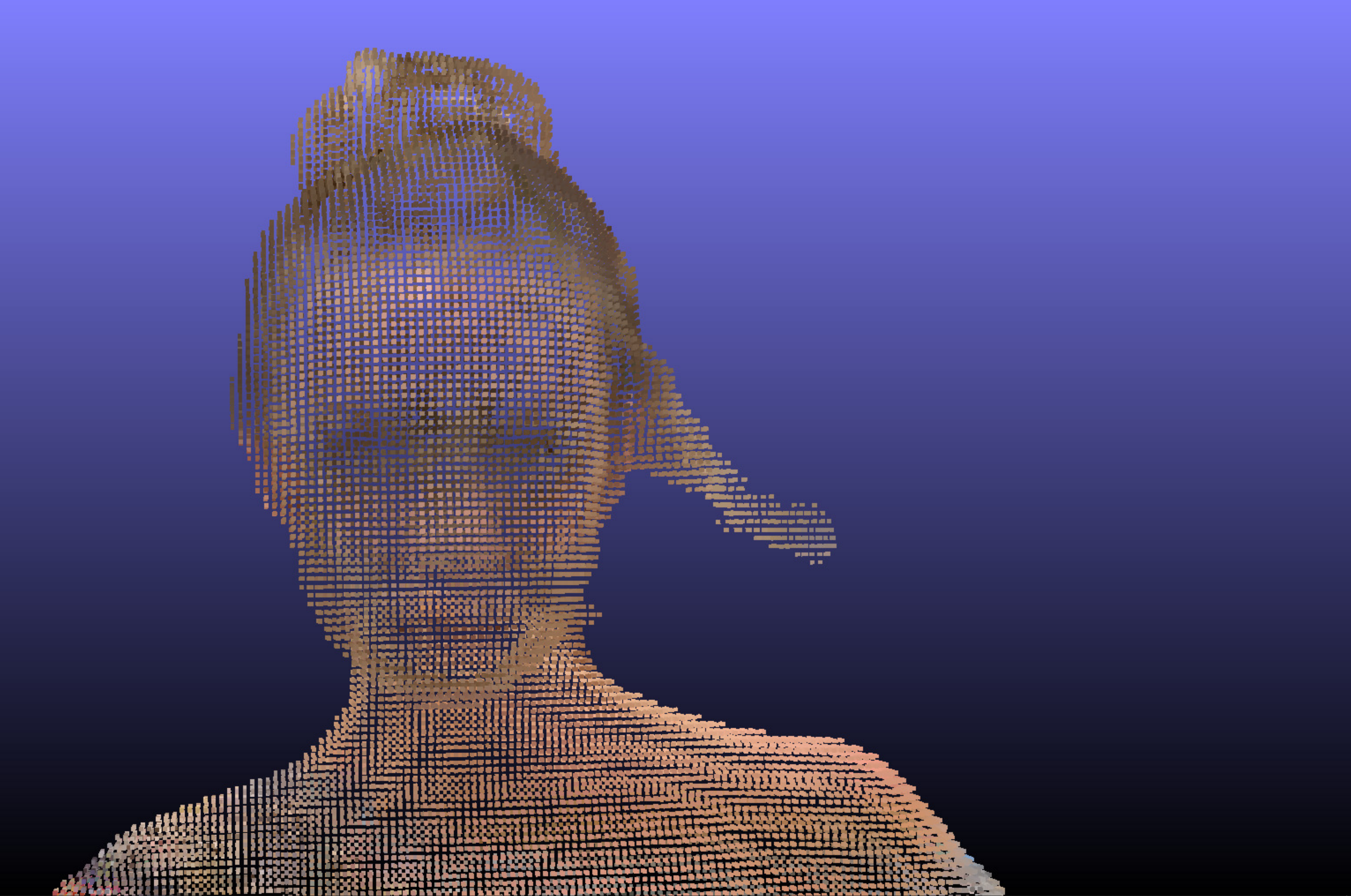}}
  \centerline{(b)}\medskip
\end{minipage}
\begin{minipage}[b]{0.48\linewidth}
  \centering
  \centerline{\includegraphics[width=0.95\linewidth]{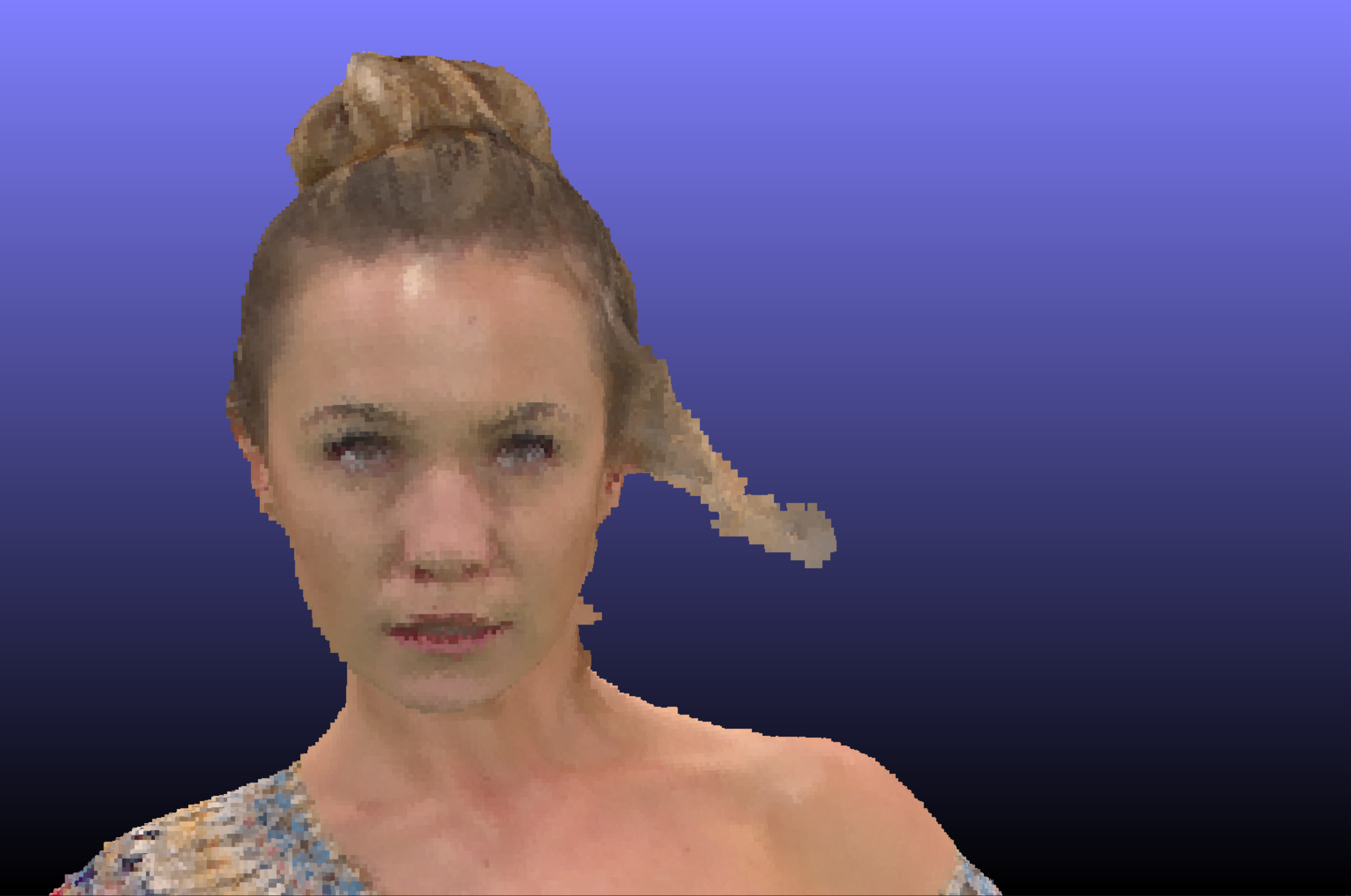}}
  \centerline{(c)}\medskip
\end{minipage}
\begin{minipage}[b]{0.48\linewidth}
  \centering
  \centerline{\includegraphics[width=0.95\linewidth]{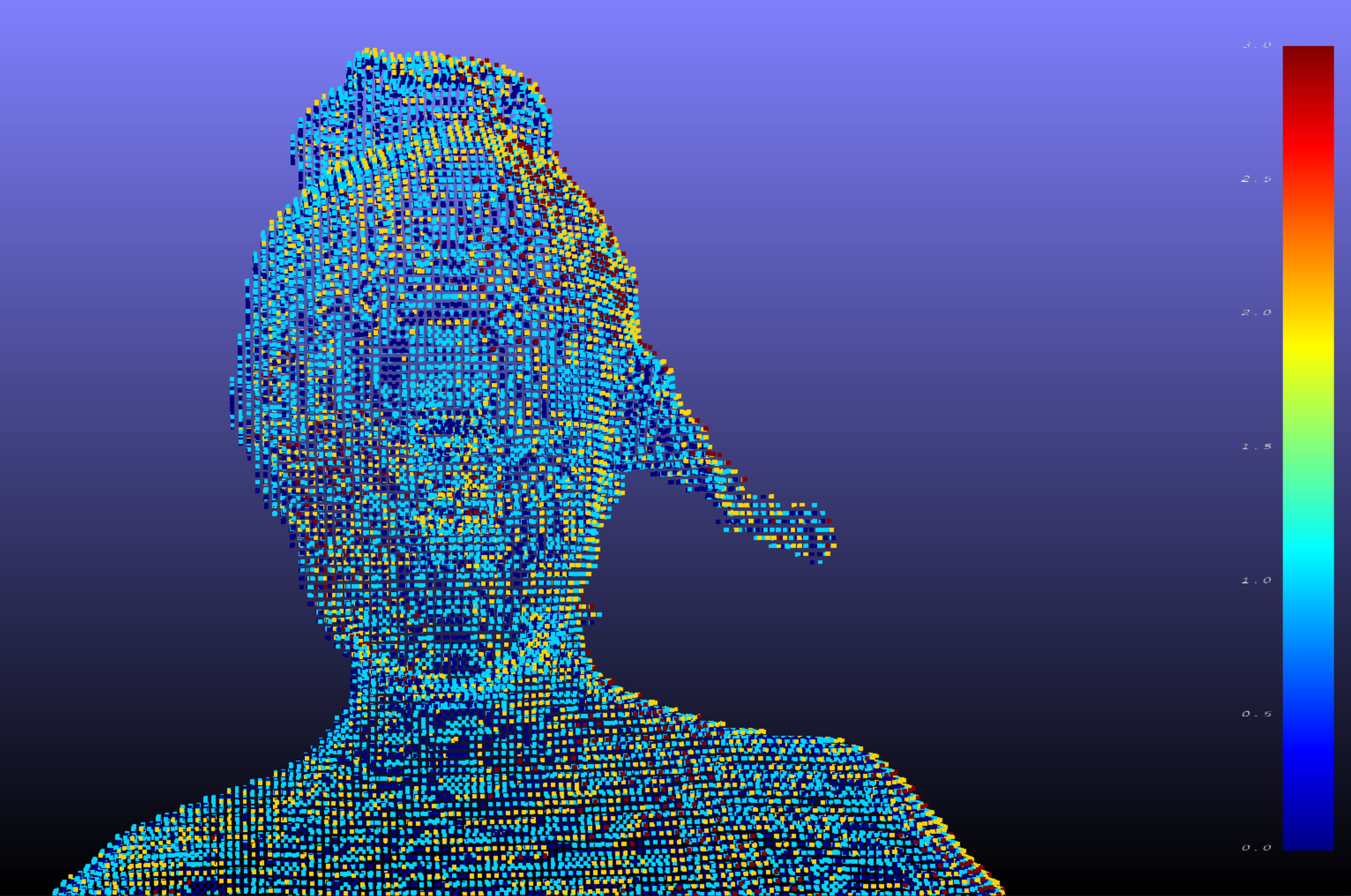}}
  \centerline{(d)}\medskip
\end{minipage}
\begin{minipage}[b]{0.48\linewidth}
  \centering
  \centerline{\includegraphics[width=0.95\linewidth]{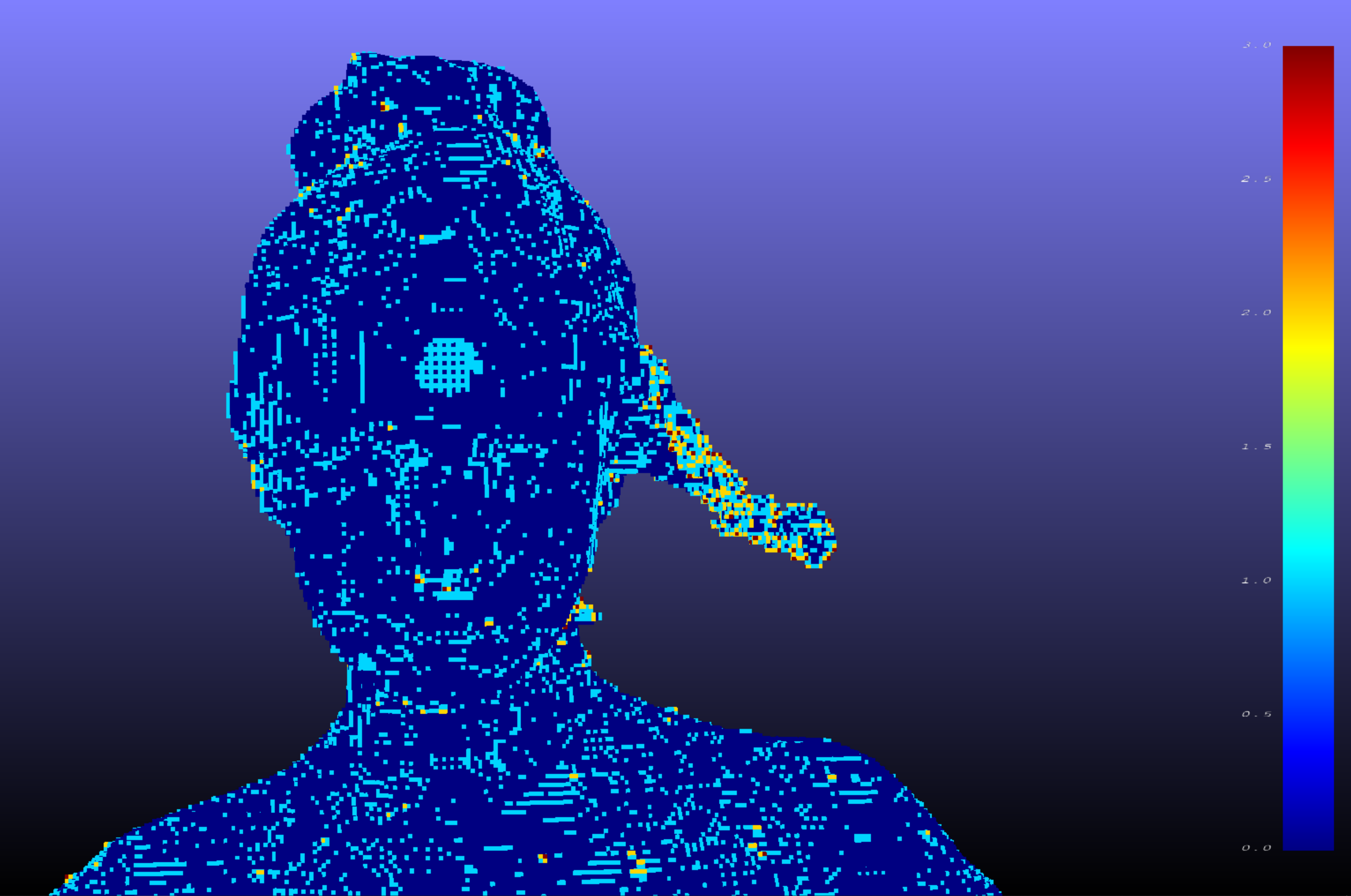}}
  \centerline{(e)}\medskip
\end{minipage}

\caption{Example of comparative results for \textit{longdress} compressed with $s=2$: (a) uncompressed ($V^{orig}$); (b)  compressed with G-PCC ($V_{d}$); (c) compressed with G-PCC+SR ($V_{sr}$); (d) $V_{d}$ \textit{vs} $V^{orig}$; (e) $V_{sr}$ \textit{vs} $V^{orig}$.}
\label{fig:longdress_prints}
\end{figure}

We also show an example of a viewpoint from \textit{longdress} for $s=2$ in Fig. \ref{fig:longdress_prints}.
Figure \ref{fig:longdress_prints}(a) shows the original uncompressed cloud.
In Figs. \ref{fig:longdress_prints}(b) and \ref{fig:longdress_prints}(c), we see the decoded point cloud under G-PCC ($V_d$) and the latter with the post-processing ($V_{sr}$), respectively. 
Important to note that the colors shown in \ref{fig:longdress_prints}(c) are obtained with a recoloring tool, only for a better observation. 
Analysing these two figures, one can clearly see the increase in the number of points and the filling of holes.
This is further emphasized in Figs. \ref{fig:longdress_prints}(d) and (e), where the Euclidean distance is shown for each voxel comparing the distorted and the original clouds, first between $V^{orig}$ and $V_d$, then between $V^{orig}$ and $V_{sr}$.
The colorbar scale ranges from 0 to 3.0. 


\section{Conclusions}
\label{sec:typestyle}

We have presented a method for post-processing point clouds which were encoded and decoded using MPEG's G-PCC codec by applying a previously proposed fractional SR technique. We compared this method with the originally decoded clouds as well as with ML-based end-to-end coding techniques. The results have shown that our method brings great improvement in quality (D1 PSNR metric) over G-PCC, specially for higher rate values (small scaling factor) in solid point clouds. In all tested cases, the method is compatible to those based on ML, except for one, which outperforms all others. However, our method does not need any sort of previous training, which is required for the ML techniques. 

\bibliographystyle{./_config/IEEEbib}
\bibliography{pointcloudrefs}

\end{document}